\documentclass[11pt]{article}
\usepackage{moriond,epsfig}

\def\Journal#1#2#3#4{{#1} {\bf #2}, #3 (#4)}

\def\NIMA{{\em Nucl. Instrum. Methods} A}
\def\NPB{{\em Nucl. Phys.} B}
\def\PLB{{\em Phys. Lett.}  B}
\def\PRL{\em Phys. Rev. Lett.}
\def\PRD{{\em Phys. Rev.} D}

\def\be{\begin{equation}}
\def\ee{\end{equation}}
\def\bea{\begin{eqnarray}}
\def\eea{\end{eqnarray}}

\begin{document}
\vspace*{4cm}
\title{SPIN EFFECTS IN FORWARD $\pi^0$-PRODUCTION IN POLARIZED PROTON-PROTON COLLISIONS AT STAR
~\footnote{ Presented at the XXXX$^{\rm th}$ Rencontres de Moriond
on QCD and High-Energy Hadronic Interactions.}}

\author{D.A. MOROZOV for the STAR Collaboration}

\address{Institute for High Energy Physics, 1 Pobeda Street, Protvino, Russia\\
e-mail: morozov@ihep.ru\\}

\maketitle\abstracts{ We report some results in large pseudo
rapidity $\pi^0$-production in polarized proton collisions at
$\sqrt{s}=200$ GeV. The single spin asymmetry for positive Feynman
$x$ ($x_F$) is consistent with zero up to $x_F\sim0.35$, then
increases with increasing $x_F$. This behavior can be described by
phenomenological models including the Sivers effect, Collins effect
or twist-3 contributions in initial and final states. The asymmetry
is found to be zero for negative $x_F$ ($-0.6<x_F<-0.2$). It has
been observed that inclusive $p+p \to \pi^0+X$ cross sections at
$\eta=3.3$, $3.8$ and $4.0$ are consistent with next-to-leading
order perturbative QCD calculations.}

\section{Introduction}
\label{sec:intro} At present Quantum Chromodynamics (QCD) can not
explain the origin of significant transverse single spin asymmetry
($A_N$) in partonic interactions. Collinear factorized perturbative
QCD (pQCD) calculations at leading twist predict these analyzing
powers to be entirely negligible, due to chirality in the theory.
However, experimental data \cite{E704,proza,E925} shows that $A_N$
for inclusive particle production is on the order of $10\%$
independent of the center of mass energy ($\sqrt{s}$). To improve
the situation theorists develop several models in a generalized
version of the QCD factorization scheme, which allows for intrinsic
transverse motion of partons inside hadrons, and of hadrons
relatively to fragmenting partons. This adds new possibilities of
spin effects, absent for collinear configurations. Sivers
\cite{Sivers} proposed as a source of spin effects to be a flavor
dependent correlation between the proton spin (${\bf S_p}$),
momentum (${\bf P_p}$) and transverse momentum (${\bf k^\perp}$) of
the unpolarized partons inside the proton. This results in the new
polarized parton distribution function:
\begin{equation}
\label{eq:sfcn} f_{q/p^{\uparrow}}(x,{\bf k^{\perp}_q};{\bf
S_p})=\hat{f}_{q/p}(x,k^{\perp}_q)
+\frac{1}{2}\Delta^{N}f_{q/p^{\uparrow}}(x,k^{\perp}_q)\frac{{\bf
S_p}\cdot ({\bf P_p}\times{\bf k^{\perp}_q})}{|{\bf S_p}||{\bf
P_p}||{\bf k^{\perp}_q}|},
\end{equation}
where $\hat{f}_{q/p}(x,k^{\perp}_q)$ - unpolarized distribution
function, $\Delta^{N}f_{q/p^{\uparrow}}(x,k^{\perp}_q)$ - Sivers
function and $x$ is the Bjorken scaling variable. Also significant
$A_N$ could be produced by the correlation between the quark spin
(${\bf s_q}$), momentum (${\bf p_q}$) and transverse momentum (${\bf
k^\perp}$) of the pion in the final state. Such an approach has been
introduced by Collins \cite{Collins}. Then the fragmentation function of
transversely polarized quark $q$ takes the form:
\begin{equation}
\label{eq:cfcn} D_{\pi/q^{\uparrow}}(z,{\bf k^{\perp}_\pi};{\bf
s_q})=\hat{D}_{\pi/q}(z,k^{\perp}_\pi)
+\frac{1}{2}\Delta^{N}D_{\pi/q^{\uparrow}}(z,k^{\perp}_\pi)\frac{{\bf
s_q}\cdot ({\bf p_q}\times{\bf k^{\perp}_\pi})}{|{\bf p_q}\times{\bf
k^{\perp}_\pi}|},
\end{equation}
where $\hat{D}_{\pi/q}(z,k^{\perp}_\pi)$ - unpolarized fragmentation
function, $\Delta^{N}D_{\pi/q^{\uparrow}}(z,k^{\perp}_\pi)$ -
Collins function and $z$ is longitudinal component of pion momentum.
Along with Collins and Sivers mechanisms there are higher twist
effects in either initial \cite{QuiSterman} or final \cite{Koike}
state which may cause the observed analyzing powers.

The Relativistic Heavy Ion Collider (RHIC) at Brookhaven National
Lab (BNL) has begun to provide collisions of polarized protons at
highest energy $\sqrt{s}=200$ GeV. The Solenoidal Tracker at RHIC
(STAR \cite{STAR}) consists mainly of a large volume TPC, Forward
TPC, Beam Beam Counters (BBC), Endcap Electromagnetic Calorimeter
(EEMC), Barrel Electromagnetic Calorimeter (BEMC) and Forward Pion
Detector (FPD). This contribution will focus on results from FPD and
BBC, which located in the very forward region of STAR coverage. BBC
are segmented scintillator detectors surrounding the beam pipe. It
provides the minimum bias trigger, absolute luminosity and relative
luminosity for our experiment. In addition BBC coincidences are used
to suppress beam gas background. FPD is a set of eight calorimeters
of lead glass cells with size of $3.8$ $cm\times3.8$ $cm\times45$
$cm$. It provides triggering and reconstruction of neutral pions.
Four of them are left-right detectors and $7\times7$ arrays cells.
Four others are top-bottom $5\times5$ arrays and are useful for
systematics studies.
\section{Single Spin Asymmetry at STAR/FPD}
\label{sec:an} 
\begin{figure}
\epsfig{figure=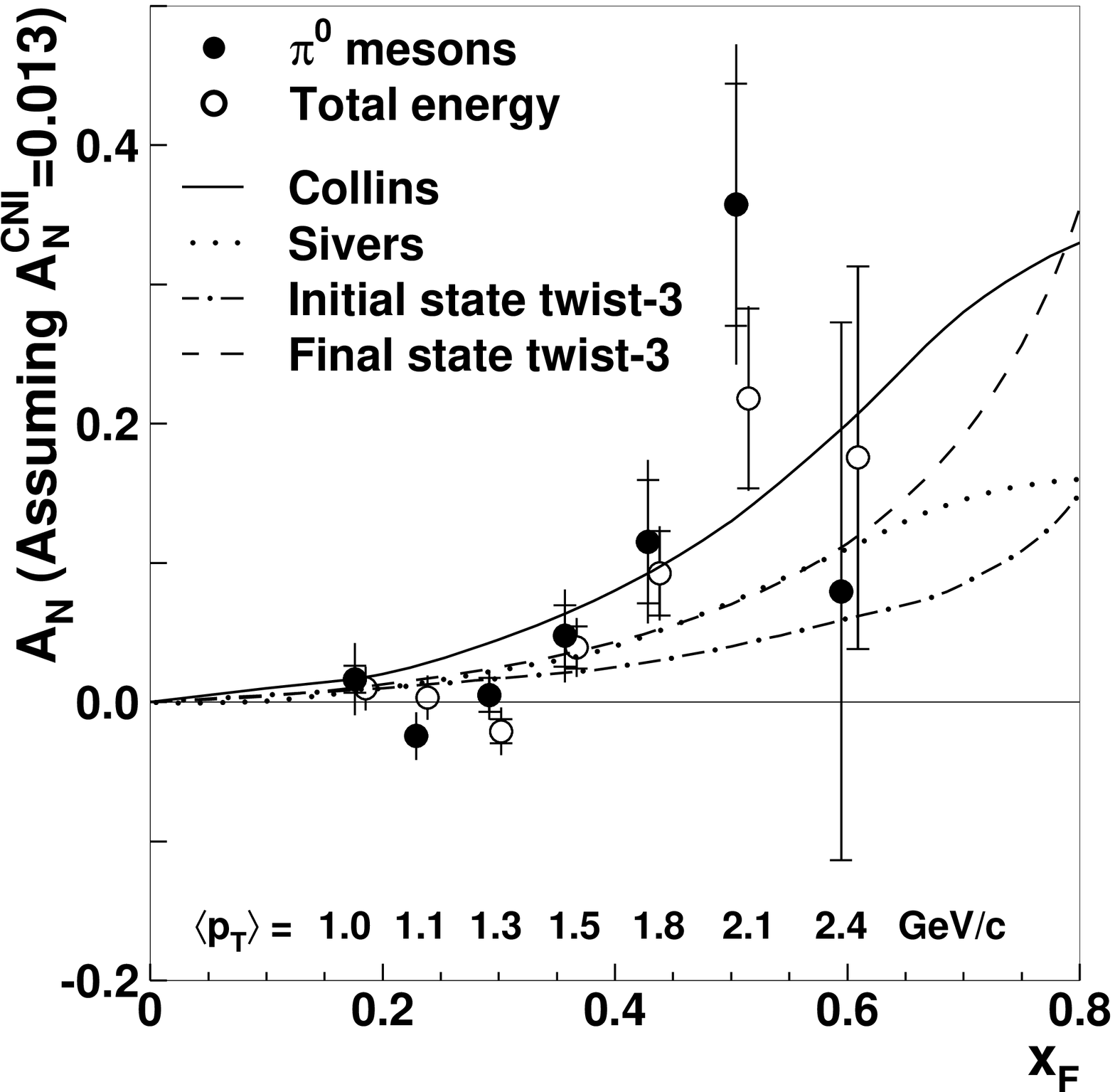,width=7.6cm}
\epsfig{figure=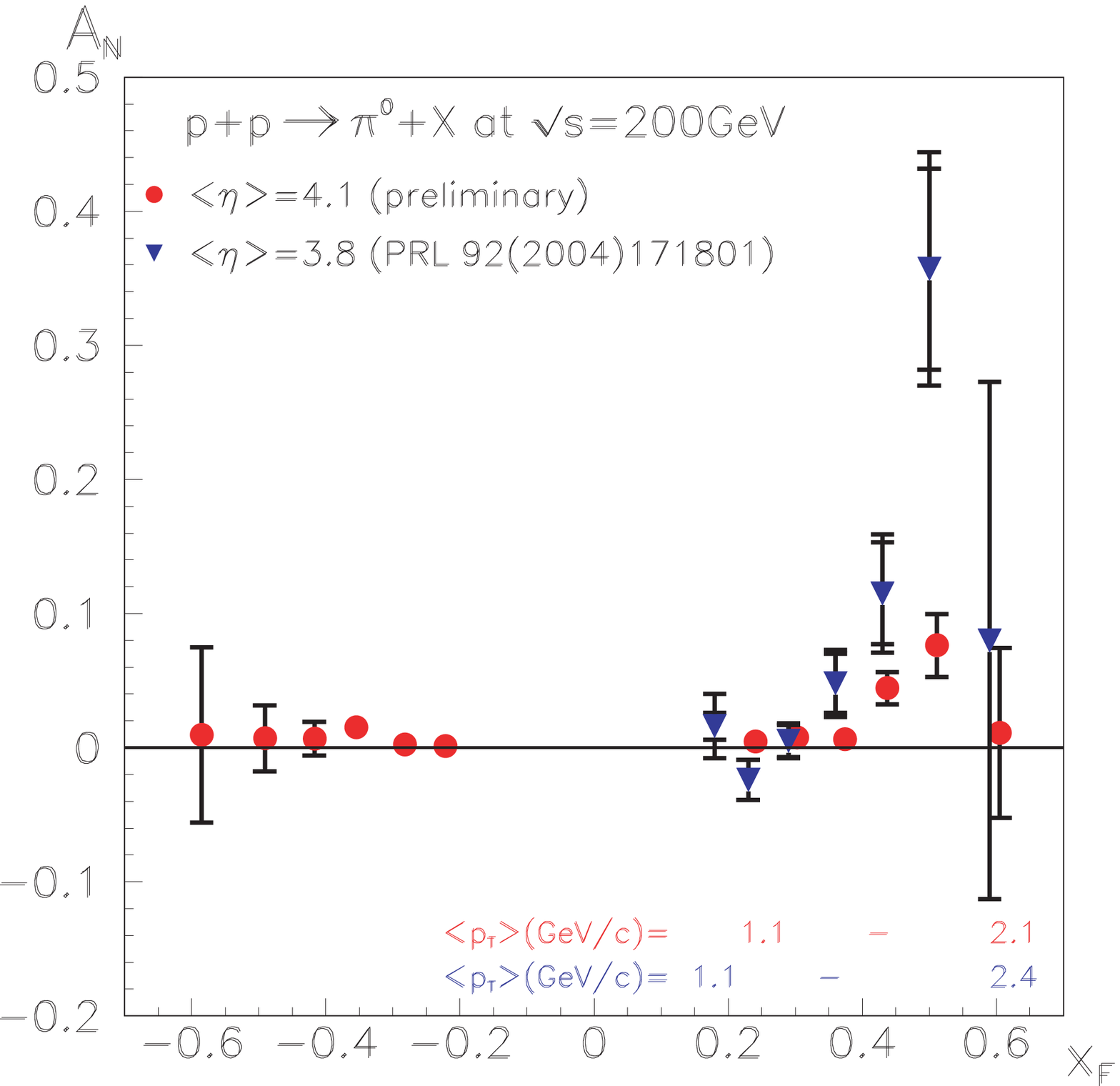,width=8.2cm} \caption{\label{fig:an}
Analyzing power of $\pi^0$ in $p^{\uparrow}p\to\pi^0X$ reaction as a
function of $x_F$. Left: for published run 2002 data at $<\eta>=3.8$
along with theoretical model curves. Right: preliminary results for
run 2003 data at $<\eta>=4.1$ (circles), compared to 2002 data
(triangles).}
\end{figure}

By definition single spin asymmetry is:
$A_N=\frac{1}{P_{Beam}}\frac{d\sigma^\uparrow-d\sigma^\downarrow}{d\sigma^\uparrow+d\sigma^\downarrow}$,
where $P_{Beam}$ - polarization of transversely polarized beam,
$d\sigma^{\uparrow(\downarrow)}$ - differential cross section of
$\pi^0$ when incoming proton has spin up(down). One can measure
$A_N$ by two different ways. First with the use of single arm
calorimeter:
$A_N=\frac{1}{P_{Beam}}\frac{N^\uparrow-RN^\downarrow}{N^\uparrow+RN^\downarrow}$,
where $N^{\uparrow(\downarrow)}$ - the number of pions detected when
the polarization of the beam is oriented up(down) and
$R=\frac{L^\uparrow}{L^\downarrow}$ is the spin dependent relative
luminosity measured by BBC. Second with the use of two arms
calorimeter (cross ratio method):
$A_N=\frac{1}{P_{Beam}}\frac{\sqrt{N^{\uparrow}_L
N^{\downarrow}_R}-\sqrt{N^{\uparrow}_R
N^{\downarrow}_L}}{\sqrt{N^{\uparrow}_L
N^{\downarrow}_R}+\sqrt{N^{\uparrow}_R N^{\downarrow}_L}}$,
where $N^{\uparrow}_{L(R)}$ - number of pions detected by the left
(right) calorimeter while the beam has spin up and
$N^{\downarrow}_{L(R)}$ - number of pions detected by the left
(right) calorimeter while the beam has spin down. In this method one
does not need the relative luminosity. The asymmetries from these
two measurements were found to be consistent and are combined.
Positive (negative) $x_F$ is defined when the pion is observed with
the same (opposite) longitudinal momentum as the polarized beam.
Positive $A_N$ is defined as more $\pi^0$ going left of the upward
polarized beam. In the 2002 proton run $0.15$ $pb^{-1}$ of
integrated luminosity was collected for transversely polarized
proton collisions at $\sqrt{s}=200$ GeV at an average polarization
of $16$$\% $. In 2003 run the integrated luminosity and average
polarization have been increased to $0.5$ $pb^{-1}$ and $25$$\% $
respectively. The polarization was measured by pC CNI polarimeter
\cite{CNI}. Fig. \ref{fig:an} shows $A_N$ versus $x_F$ for $\pi^0$
mesons.
Left plot represents published 2002 data for $<\eta>=3.8$
\cite{fpdprl}. Result is consistent with measurements at lower
$\sqrt{s}=20$ GeV (E704 experiment) and increasing with $x_F$. It
also can be described by all theoretical predictions mentioned above
due to statistical uncertainties. Results from run 2002 and
preliminary results from run 2003 at $<\eta>=4.1$ \cite{Ogawa} are
compared on the right plot. The analyzing power for positive $x_F$
at $<\eta>=4.1$ is consistent with zero up to $x_F\sim0.35$, then
increases with increasing $x_F$. The first measurement of $A_N$ at
negative $x_F$ has been done, and is found to be zero. Negative
$x_F$ results may give an upper limit on the gluon Sivers function
\cite{Anselmino}. This work is in progress.

\section{Differential cross sections for forward $\pi^0$-Production}
\label{sec:cros} 
\begin{figure}
\epsfig{figure=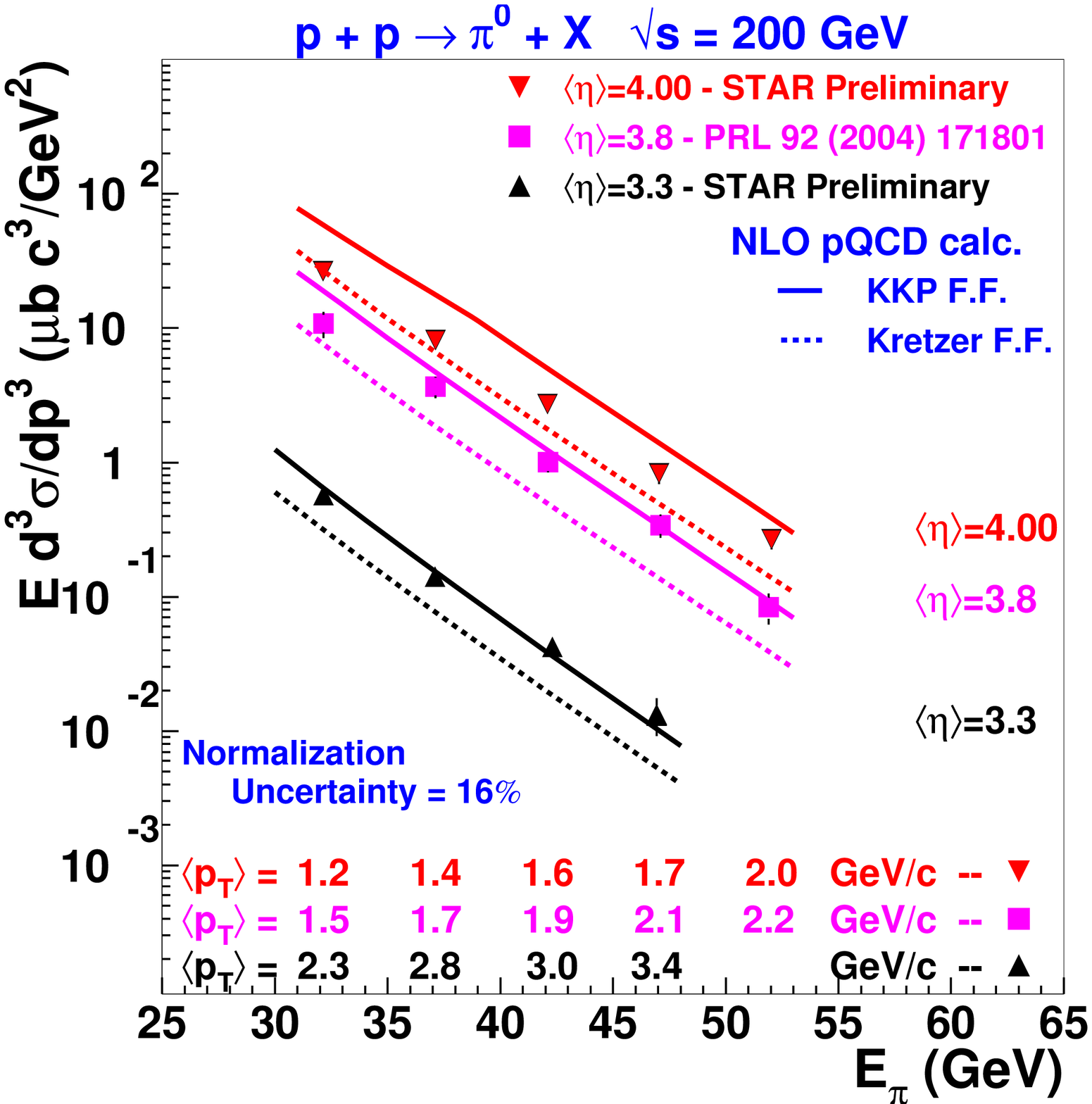,width=8.0cm}
\epsfig{figure=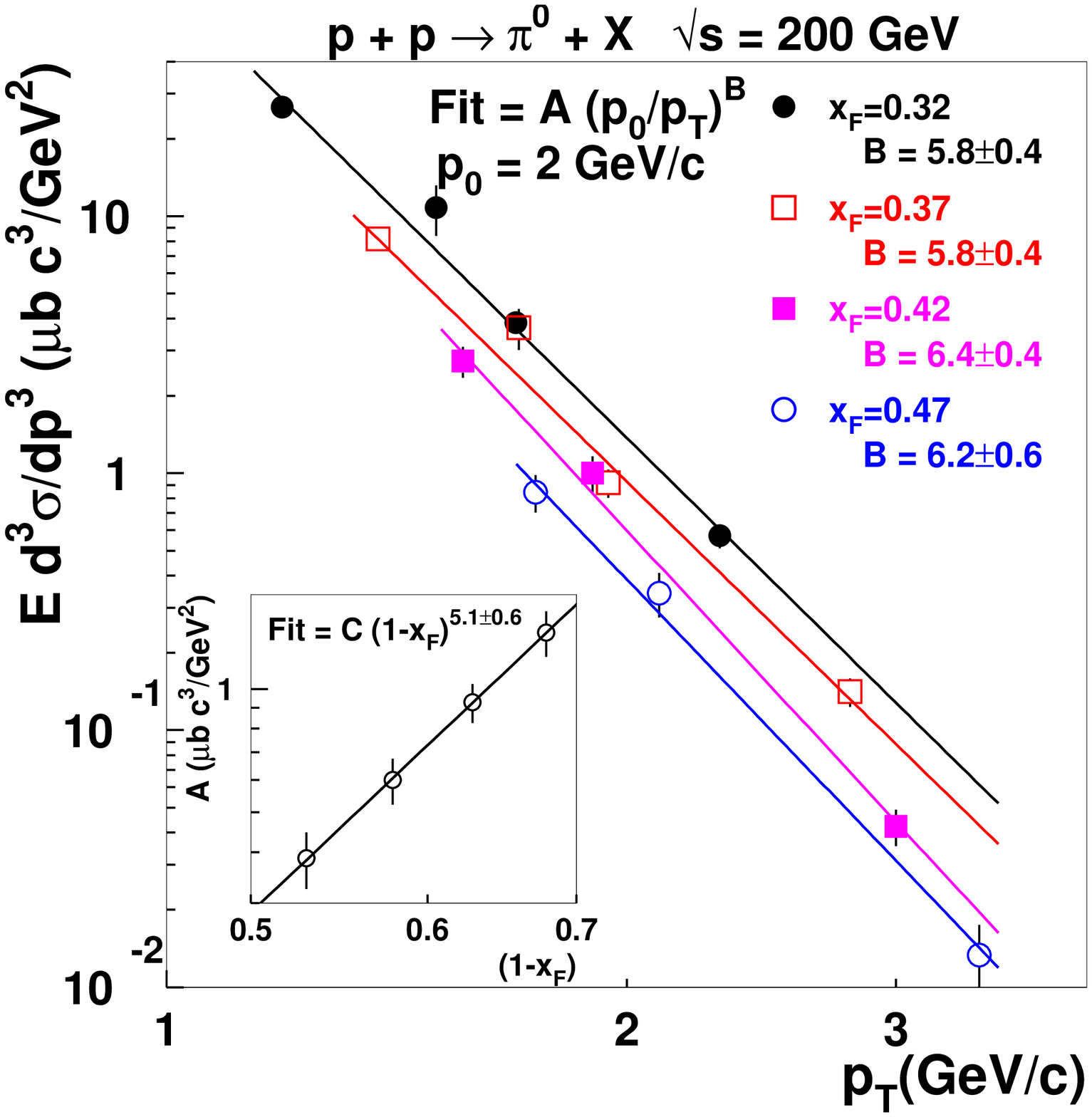,width=7.8cm}
\caption{\label{fig:cs}
Left: invariant cross section for $\pi^0$ produced in $pp$
collisions at $\sqrt{s}=200$ GeV versus pion energy ($E_\pi$) at
average pseudorapidities ($<\eta>$) $3.3$, $3.8$ and $4.0$. The
error bars are point-to-point systematic and statistical errors
added in quadrature. Right: invariant cross section as a function of
$p_T$ at fixed $x_F$ (outer) and as a function of $(1-x_F)$ at fixed
$p_T=2$ GeV/c (inner). Lines are fits by the functions showed in the
plot.}
\end{figure}

The inclusive differential cross section for
$\pi^0$ production for $30<E_\pi<55$ GeV at $<\eta>=3.8$ was
previously published \cite{fpdprl}. The result at $<\eta>=3.3$ in
2002 run also have been extracted. In 2003 run new calorimeters and
readout electronics have been installed to allow measurements of the
differential cross section at $<\eta>=4.0$. The results are shown in
Fig. \ref{fig:cs}.

On the left plot the cross sections are shown versus pion energy and
are compared with NLO pQCD calculations evaluated at $\eta=3.3$,
$3.8$ and $4.0$. Two sets of fragmentation functions are used. The
model calculations are consistent with the data in contrast to the
data at lower $\sqrt{s}$ (NLO pQCD calculations at $\sqrt{s}=20$ GeV
underpredict measured cross sections \cite{BorSof}). As $\eta$
increases, systematics regarding the comparison with NLO pQCD
calculations begin to emerge. The data at low $p_T$ are more
consistent with the Kretzer set of fragmentation functions. Similar
trend was observed at mid-rapidity \cite{phenixcs}. On the right
plot the data is represented as in earlier experiments \cite{ISR}.
The outer picture shows cross section as a function of $p_T$ at
fixed $x_F$. The inner one -- cross section as a function of
$(1-x_F)$ at fixed $p_T=2$ GeV/c. Invariant cross section falls with
$p_T$ at fixed $x_F$ with exponent (value $\sim6$) independently on
$x_F$. Data also show exponential dependence on $x_F$ with fixed
$p_T=2$ GeV/c. The value of the fitted exponent ($\sim5$) may be
sensitive to the interplay between hard and soft scattering
processes. One note should be stated regarding systematics of this
separated $x_F$ and $p_T$ dependencies. Data were accumulated in
different conditions in different running years: with different
calorimeters, with different readout electronics, taken in different
kinematical regions.
\section{Summary}
\label{sec:sum}
Large spin effects have been observed at forward $\pi^0$ production in polarized
$pp$ collisions at highest energy $\sqrt{s}=200$ GeV at STAR FPD. The single spin
asymmetry for positive $x_F$ is consistent with zero up to $x_F=0.35$, then
increases with increasing $x_F$. The asymmetry is found to be zero for negative
$x_F$. The inclusive differential cross section for forward $\pi^0$ production at
$\sqrt{s}=200$ GeV is consistent with NLO pQCD calculations in contrast to what was
observed at lower energy. First try to map the cross section in $x_F-p_T$ plane
was performed.

The near-future plans are to increase the statistics with the
present FPD in order to measure the $p_T$ dependence of $A_N$ at
fixed $x_F$, and to extract the gluon Sivers function.  In the
longer term, plans include an increase of the angular coverage of
the electromagnetic calorimetry in the forward direction, with a
goal to disentangle the dynamical origin of transverse single spin
asymmetries \cite{FMS}.

\end{document}